\documentstyle[twocolumn,aps,prl,psfig]{revtex}

\draft

\begin{document}

\title{A Lattice Boltzmann Model for Wave and Fracture phenomena}

\author{Bastien Chopard, Pascal Luthi and St\'ephane Marconi}

\address{
Computer Science Department, University of Geneva\\
CH 1211 Gen\`eve 4, Switzerland}

\date{\today}

\maketitle

\begin{abstract} 
We show that the lattice Boltzmann formalism can be used to describe
wave propagation in a heterogeneous media, as well as solid-body-like
systems and fracture propagation.  Several fundamental properties of
real fractures (such as propagation speed and transition between rough
and smooth crack surfaces) are well captured by our approach.
\end{abstract}


Lattice Boltzmann (LB)
models~\cite{quian-succi:96,BC-livre} are dynamical
systems, discrete in time and space, aimed at simulating the behavior
of a real physical system in terms of local density of fictitious
particles moving and interacting on a regular lattice. The density
distribution functions are denoted $f_i(\vec r,t)$ where $\vec r$ refers to
the lattice site, $t$ the iteration time while the subscript $i$
labels the admissible speed of motion $\vec v_i$ (e.g.  along the main
lattice directions). The value $i=0$ corresponds to a population of
rest particles with $\vec v_0=0$.

Lattice BGK models~\cite{qian:92} have been used successfully to simulate
fluid dynamics and complex flows~\cite{quian-succi:96,BC-livre}.
The same approach can be adapted to model wave propagation in a heterogeneous
media, where propagation speed, absorption and reflection can be adjusted
locally for each lattice sites. We show how such a model can be derived and
applied to the study of fracture propagation

Lattice BGK models are characterized by the following dynamics
\begin{equation}
f_i(\vec r+\tau\vec v_i,t+\tau)-f_i(\vec r,t)=
{1\over\xi}\left(f_i^{(0)}(\vec r,t)-f_i(\vec r,t)\right)
\label{lbe-wave-1}
\end{equation}
where $\tau$ is the time step, $f_i^{(0)}(\vec r,t)$ the so-called local
equilibrium distribution and $\xi$ a relaxation time.  The function
$f_i^{(0)}$ is the key ingredient for it actually contains the properties
of the physical process under study: this is the distribution to which
the dynamics spontaneously relaxes and which is, therefore, intimately
related to the nature of the system.

Wave phenomena, whether mechanical or electromagnetic derives from two
conserved quantities $\Psi$ and $\vec J$, together with time reversal invariance
and a linear response of the media. The quantity
$\Psi$ is a scalar field and
$\vec J$ its associated current. For sound waves, $\Psi$ and $\vec J$ are
respectively the density and the momentum variations. In electrodynamics,
$\Psi$ is the energy density and $\vec J$ the Poynting vector~\cite{jackson}.

The idea behind the LB approach is to ``generalize'' a physical process to
a discrete space and time universe, so that it can be efficiently simulated
on a (parallel) computer. For waves, this generalization is obtained by
keeping the essential ingredients of real phenomena, namely conservation of
$\Psi$ and $\vec J$, linearity and time reversal invariance. Thus, in a
discrete space-time universe, a generic system leading to wave propagation
is obtained from eq.~(\ref{lbe-wave-1}) by an appropriate choice of the
local equilibrium distribution
\begin{eqnarray}
f_i^{(0)}=a\Psi + b{\vec v_i\cdot\vec J \over 2v^2} \qquad{\rm if\ }
i\ne 0,\ \rm and\ \qquad f_0^{(0)}=a_0\Psi
\label{leq-conditions} 
\end{eqnarray}
where $v$ is the ratio of the lattice spacing to the time step, and $\Psi$
and $\vec J$ are related to the $f_i$s in the standard way: $\Psi=\sum_{i} f_i$
and $\vec J=\sum_i f_i\vec v_i$. Note that, here, we make no restriction on the sign
of the $f_i$s which may well be negative in order to represent a wave.

As opposed to hydrodynamics~\cite{quian-succi:96}, $f_i^{(0)}$ is a
linear function of the conserved quantities, which ensures the
superposition principle.  The parameters $a$, $b$ and $a_0$ are chosen
so that $\Psi=\sum_{i}f_i^{(0)}$ and $\vec J=\sum_{i} \vec v_i f_i^{(0)}$,
which ensures conservation of $\Psi$ and $\vec J$.  For a two-dimensional
square lattice (D2Q5 to use the terminology of~\cite{quian-succi:96})
we find $a_0 + 4a=1$ and $b=1$.  The freedom on the value of $a_0$ can
be used to adjust locally the wave propagation speed. Time reversal
invariance is enforced by choosing $\xi=1/2$ as can be easily
checked from eq.~(\ref{lbe-wave-1}) with $\vec J\to-\vec J$ and $\Psi\to\Psi$
in relation~(\ref{leq-conditions}).  Note that the D2Q5 lattice is
known for giving anisotropic contributions to the hydrodynamic
equations. These terms are not present in our wave model because they
appear with a vanishing coefficient when $\xi=1/2$ (see
eq.~(\ref{wave-mvt-2}).

In hydrodynamic models, $\xi=1/2$ corresponds to the limit of zero
viscosity, which is numerically unstable. In our case, this instability
does not show up provided we use an appropriate lattice. Indeed, in the
D2Q5 lattice, our dynamics is also unitary~\cite{luthi-phd:98} which
ensures that $\sum_{i}f_i^2$ is conserved. This extra condition prevents
the $f_i$s from becoming arbitrarily large (with positive and negative
signs, since $\Psi$ is conserved). This is no longer the case with the D2Q9
lattice, where numerical instabilities develop for our wave dynamics.
This observation may shed some light on the origin of the numerical
instabilities observed in hydrodynamic models.

Note that dissipation can be included in our microdynamics. Using
$\xi>1/2$ allows us to describe waves with viscous-like
dissipation. This makes sense with the hexagonal lattice D2Q7, where
no stability problem occurs when $\xi=1/2$ and no anisotropy problem
appears when the viscosity is non-zero ($\xi>1/2$). Below we shall propose
another way to include dissipation in the square lattice model, which
will be appropriate to our purpose of modeling fracture propagation.

The multiscale Chapman-Enskog expansion~\cite{BC-livre} can be used to
derive the macroscopic behavior of $\Psi$ and $\vec J$ when the lattice
spacing and time step go to zero. We obtain
\begin{equation}
\partial_t\Psi + \partial_\beta J_\beta=0 \label{wave-cont}
\end{equation}
\begin{eqnarray}
\partial_t J_\alpha &+&2av^2\partial_\alpha\Psi  + \nonumber\\
& & \left(2\xi-1\right) \left[a\tau v^2\partial_\alpha{\rm div}\vec J 
  -{\tau\over4v^2}
         T_{\alpha\beta\gamma\delta}\partial_\beta\partial_\gamma J_\delta \right]=0
\nonumber\\ \label{wave-mvt-2}
\end{eqnarray}
where $T_{\alpha\beta\gamma\delta}=\sum_i v_{i\alpha}v_{i\beta}v_{i\gamma}v_{i\delta}$
and summation over repeated greek
indices (which label the spatial coordinates) is assumed. With $\xi=1/2$
equation~(\ref{wave-mvt-2}) becomes 
$\partial_t J_\alpha +2av^2\partial_\alpha\Psi= 0$. When combined with
equ.~(\ref{wave-cont}), we obtain
\begin{eqnarray}
\partial^2_t\Psi -2av^2\nabla^2\Psi&=&0 \nonumber
\end{eqnarray}
which is a wave equation with propagation speed $c=v\sqrt{2a}$ (note
that $v$ is the speed at which information travels). As mentioned
previously, the propagation speed $c$ can be adjusted from place to
place by choosing the spatial dependency of $a$. Provided that
$a_0+4a=1$ and $a_0\ge0$ (required for stability reasons), the largest
possible value is $a=1/4$ and corresponds to a maximum velocity
$c_0=v/\sqrt{2}$.  Therefore media with different refraction indices
$n=c_0/c=1/(2\sqrt{a})$ can be modeled.

Perfect reflection on obstacles can be included by modifying the
microdynamics to be $f_i(\vec r+\tau\vec v_i,t+\tau)=-f_{i'}(\vec
r,t)$ on mirror sites, where $i'$ is defined so that $\vec
v_{i'}=-\vec v_i$, i.e. the flux bounces back to where they came from
with a change of sign. Absorption on non-perfect transmitter sites can
be obtained by modifying the conservation of $\Psi$ to
$\sum_{i}f_i^{(0)}=\mu\Psi$, where $0\le\mu\le1$ is an attenuation
factor. This modifies $a\to\mu a$ and $a_0\to\mu a_0$. Finally, by
substituting~(\ref{leq-conditions}) into (\ref{lbe-wave-1}) and using
the expression of $a$ and $a_0$ in terms of $c$, free propagation with
refraction index $n(\vec r)$, and partial transmission and reflection
can be expressed as
\begin{eqnarray}
f_i(\vec r+\tau\vec v_i,t+\tau)&=&{\mu\over2n^2}\Psi-f_{i+2}(\vec r,t) 
\nonumber\\
f_0(\vec r, t+\tau)&=&2\mu{n^2-1\over n^2}\Psi-f_0(\vec r,t)
\label{lb-wave}
\end{eqnarray}
In this equation, $\mu=0$ corresponds to perfect reflection, $\mu=1$
to perfect transmission and $0<\mu<1$ describes a situation where the
wave is partially absorbed.  A particular version of our LB wave model
has been successfully validated by the problem of radio wave
propagation in a city~\cite{BC-iee}.

The idea of expressing wave propagation as a discrete formulation of
the Huygens principle has been considered by several
authors~\cite{hof:85,hrg:92,van:92,vanneste:97}. Not surprisingly, the
resulting numerical schemes bear a strong similarity to
ours. Nevertheless the context of these studies is different from ours
and none have noticed the existing link with the lattice BGK
approach. Models of refs.~\cite{van:92,vanneste:97} use a reduced set
of conserved quantities, which may not be appropriate in our
case. Other models~\cite{mora:92} consider wave
propagation in a LB approach, but with a significantly more
complicated microdynamics and a different purpose.

In what follows, we show how our LB dynamics can model a solid body and
capture the generic feature of crack propagation.  Whereas LB methods have
been largely used to simulate systems of point particles which interact
locally, modeling a solid body with this approach (i.e modeling an object
made of many particles that maintains its shape and coherence over
distances much larger than the interparticle spacing) has remained mostly
unexplored.  A successful attempt to model a {\em one-dimensional} solid as
a cellular automata is described in~\cite{BC-string}. The crucial
ingredient of this model is the fact that collective motion is achieved
because the ``atoms'' making up the solid vibrate in a coherent way and
produce an overall displacement. This vibration propagates as a wave
throughout the solid and reflects at the boundary.

A 2D solid-body can be thought of as a square lattice of particles linked
to their nearest neighbors with a spring-like interaction. Generalizing the
model given in \cite{BC-string} requires us to consider this solid as made
up of two sublattices. We term them black and white, by analogy to the
checkerboard decomposition. The dynamics consists in moving the black
particles as a function of the positions of their white, motionless
neighbors, and vice-versa, at every other steps.

\begin{figure}
\centerline{\psfig{figure=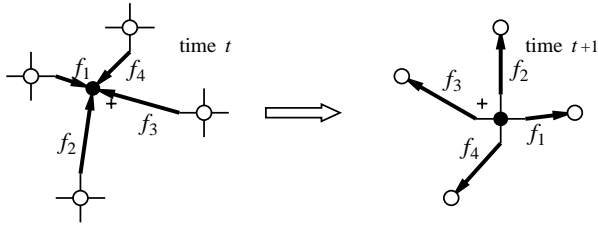,width=8cm}}
\caption{Illustration of the way the $\vec f_i$s are defined. The cross
indicates the location of the geometrical center of mass of the four white
particles. At the next iteration, the black particle jumps to a symmetrical
position with respect to this point.}
\label{fig:solid-2d}
\end{figure}

Let us denote  the location of a black particle by $\vec r_{i,j}=(x_{i,j},
y_{i,j})$. The surrounding white particles will be at positions $\vec r_{i-1,j}$,
$\vec r_{i+1,j}$, $\vec r_{i,j-1}$ and $\vec r_{i,j+1}$. We define the
separation to the central black particle as (see figure~\ref{fig:solid-2d})
\begin{eqnarray}
\vec f_1(i,j)&=&\vec r_{i,j}-(\vec r_{i-1,j}+\vec h) \nonumber \\
\vec f_2(i,j)&=&\vec r_{i,j}-(\vec r_{i,j-1}+\vec u) \nonumber \\
\vec f_3(i,j)&=&\vec r_{i,j}-(\vec r_{i+1,j}-\vec h) \nonumber \\
\vec f_4(i,j)&=&\vec r_{i,j}-(\vec r_{i,j+1}- \vec u) \nonumber \\
\label{f-to-x}
\end{eqnarray}
where the $\vec f_i$ are now vector quantities, and $\vec h=(r_0,0)$
and $\vec u=(0,r_0)$ can be thought of as representing some equilibrium
length $r_0$ of the horizontal and vertical spring connecting
adjacent particles.  With this formulation, the coupling between
adjacent particles is not given by the Euclidean distance but is
decoupled along each coordinate axis (however, a deformation along the
$x$-direction will propagate along the $y$-direction and conversely).
This method makes it possible to work with a square lattice, which is
usually not taken into account when describing deformation in a solid
because, with the Euclidean distance, the $y$-axis can be tilted by an
angle $\alpha$ without applying any force. The breaking of the
rotational invariance is expected not to play a role in the fracture
process we shall consider below.

The motion of all black particles is obtained by updating the above $\vec
f_i$s by equ.~(\ref{lb-wave}), with $n=1$ and for $i>0$. The local value of
$\vec \Psi$ (which is conserved by the dynamics) is then interpreted as the
momentum $\vec p$ of the corresponding particle. The new location $\vec
r_{ij}(t+1)$ of particle $ij$ is thus obtained as $\vec r_{ij}(t+1)=\vec
r_{ij}(t)-(1/m_{ij})\vec\Psi$, where $m_{ij}$ is a mass associated with the
particle. Next, the quantities $\vec f$ are propagated to the neighbors
and interpreted as the deformations seen by the white particles,
which then move according to the same procedure.

A geometrical interpretation of this dynamics is given in
fig.~\ref{fig:solid-2d}: for $n=1$, a bulk particle moves to a symmetric
location with respect to the center of mass of its neighbors, $(1/4)[\vec
r_{i-1,j}+\vec h+\vec r_{i+1,j}-\vec h+\vec r_{i,j-1}+\vec u+\vec
r_{i,j+1}-\vec u]$. Thus, our model corresponds to looking at particles in
a harmonic potential, at the particular time where they have maximal
displacement and no velocity. Note also that, in this interpretation $\vec J_x$
and $\vec J_y$ give, respectively, the separation between the left-right
and up-down neighbors.

At the boundary of the domain, or for particles with broken bonds, a
different rule of motion has to be considered. The above
interpretation, in terms of a jump with respect to the center of mass
of the neighbors, gives a natural way to formulate the dynamics when
less than four links are present.

Our original wave paradigm includes five fields. In the case of a solid
body, the fifth quantity $f_0$ can be added in the model as an internal
degree of freedom. This is useful in order to describe solids with
different sound speeds $c$ and to then see how the fracture propagation
speed relates to $c$. The particle motion is still determined by the local
momentum $\vec \Psi$ but, now, there is no more distinction between white
and black particles.

Our goal is now to show that our LB wave model can be used to describe a
fracture process. Fracture is a phenomena for which no definite theory is
available~\cite{marder:96} and a simple model is certainly
useful to help understanding generic properties.

At the level of our description, a fracture is easily introduced. A
link may break locally if its deformation is too large. Here we
consider the energy stored in a link as the quantity determining the
breaking. The energy $E_k(\vec r,t)$ of each link $k$, ($k=1$ to $4$)
at site $\vec r$ and time $t$ is defined as $E_k=(1/4)\vec f_0^2 +
\vec f_k^2$. Since, as mentioned above, our dynamics is unitary, the
total energy $E_{{\rm tot}}=\sum_{k,\vec r} E_k(\vec r)$ is conserved
until a link breaks. Note that, according to our interpretation, the
microscopic energy is only of potential type at integer time steps.

The breaking rule we impose is as follows:  a link $k$ breaks if the
corresponding $E_k$ is larger than a given threshold $\epsilon(\vec r)$ which
may, in principle, depend on the position (local defects). Particles with
one or more broken links then behave like particles at the boundary.

The total energy $E_{{\rm tot}}$ can be written as the kinetic energy
of the center of mass $E_{{\rm kin}}$, plus an internal energy
$E_{{\rm int}}\equiv E_{{\rm tot}}-E_{{\rm kin}}$. The kinetic part is
computed as $E_{{\rm kin}}=(1/2M)[\sum \vec p_{ij}]^2$, where $M$ is the
total mass.

The second contribution, $E_{{\rm int}}$, can be set
proportional to a temperature $T$, using the equipartition theorem. In
the initial configuration, $T$ is typically introduced by adding a
noise of standard deviation $\sqrt{T}$ to the rest position of each
atom.

\begin{figure}
\centerline{\psfig{figure=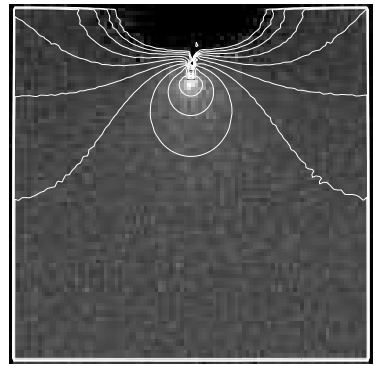,width=5cm}}
\caption{Contour plot of the energy $E$ in a $100\times 100$ square solid 
under a preloaded (mode I) stress ($S=2.2\times 10^{-2}$). The contour
line mark the levels $\epsilon/E=10\times\ell$, with $\ell=1,...,10$.}
\label{contrainte}
\end{figure}

A typical experiment which is performed when studying fracture
formation is to apply a stress by pulling in opposite directions the
left and right extremities of a solid sample. To achieve a static
stress, the solid is prepared in a configuration where the $x$-spacing
between the atoms is increased to the value $(1+S)r_0$ where $S$ is
called the stress factor. Both left and right extremities are not
allowed to move. The initial temperature may be different from 0 and a
small notch (artificially broken links) is made in the middle of the
sample to favor the apparition of the fracture at this position. In
the fracture community, this is known as mode I
loading~\cite{herrmann-roux:90}.

\begin{figure}
\centerline{\psfig{figure=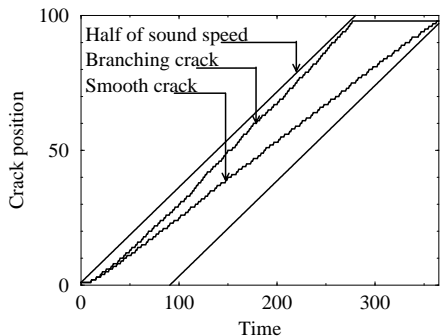,width=6cm}}
\caption{Fracture tip location  as a function of time (iterations) for two
types of cracks; for comparison, straight lines corresponding to half the
speed of sound  are drawn.  The fastest crack corresponds to a branching
situation (see figure~\ref{crack}), obtained for $S=0.03$,
$\mu=0.96$, $\epsilon=0.0058$. The slowest crack is smooth (see
figure~\ref{crack}) and is produced with $S=0.03$, $\mu=0.91$,
$\epsilon=0.0058$.}
\label{crackspeed}
\end{figure}

Figure~\ref{contrainte} shows the stationary spatial distribution of stress
around the notch given by the energy contour lines and obtained from our
simulation in a case where the fracture does not propagate across the
sample but stops after a few steps.  In figure~\ref{crackspeed} we can see
the result of two fracture experiments where the crack spontaneously
propagates through the sample after being initiated artificially. The
dissipation coefficient $\mu$ in~(\ref{lb-wave}) turns out to be essential
and eventually distinguishes the two cases presented here. Attenuation
prevents the reflection of too much energy from the boundaries toward the
crack.  With $\mu=0.91$ the dissipation of energy is high enough to limit
the acceleration of the crack below some critical speed and the crack
remains smooth. On the other hand, with $\mu=0.96$ the crack accelerates
above this critical speed and instabilities appear: the crack progresses
while making micro-branching. 

\begin{figure}
\centerline{\psfig{figure=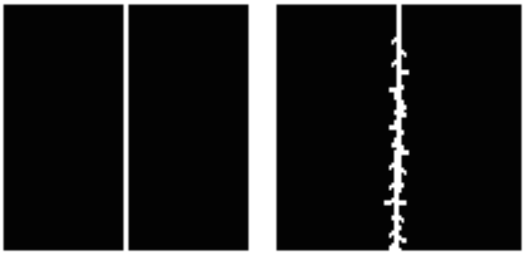,height=3.5cm}}
\caption{Map of the broken links for the smooth (left) and branching cracks
(right) described in figure~\ref{crackspeed}.}
\label{crack}
\end{figure}

As can be seen in figure~\ref{crackspeed}, the limiting speed is
around half of the speed of sound in the sample.  The simulations were
performed on samples with $c=1/\sqrt{2}$, but we observe a similar
behavior using another values of $c$. Thus, a critical fracture
speed~\cite{marder:96} of about $c/2$ for the smooth--branching
transition is well captured in our model. This non-trivial result is
promising since no simple statistical models are yet available to
describe a fracture process. The fact that our dynamics is based on a
description at the particle level makes the connection with real
experiment possible and leaves much flexibility to adjust locally some
parameters. Yet, our model is significantly simpler than a molecular
dynamics approach.  As quantitative experimental results are now
possible at a microscopic scale (e.g. scaling in the apparition of
micro-cracks before the main crack~\cite{ciliberto:97}), our model
could contribute to highlight the essential mechanisms of fracture.
Another interesting check would be to measure, both numerically and
experimentaly the relation between roughness and fracture speed.

We acknowledge support from the Swiss National Science Foundation.


\begin{thebibliography}{10}

\bibitem{quian-succi:96}
Y.H. Qian, S.~Succi, and S.A. Orszag.
\newblock In D.~Stauffer, editor, {\em Annual Reviews of Computational Physics
  III}, pages 195--242. World Scientific, 1996.

\bibitem{BC-livre}
B.~Chopard and M.~Droz.
\newblock {\em Cellular Automata Modeling of Physical Systems}.
\newblock Cambridge University Press, 1998.

\bibitem{qian:92}
Y.H. Qian, D.~d'Humi\`eres, and P.~Lallemand.
\newblock {\em Europhys. Lett}, 17(6):470--84, 1992.

\bibitem{jackson}
J.D. Jackson.
\newblock {\em Classical Electrodynamics}.
\newblock John Wiley, 1975.

\bibitem{luthi-phd:98}
P.O. Luthi.
\newblock PhD thesis, Computer Science Department, University of Geneva, 24 rue
  General-Dufour, 1211 Geneva 4, Switzerland, 1998.

\bibitem{BC-iee}
B.~Chopard, P.O. Luthi, and {Jean-Frédéric} Wagen.
\newblock {\em IEE Proceedings - Microwaves, Antennas and Propagation},
  144:251--255, 1997.

\bibitem{hof:85}
W.~J.~R. Hoeffer.
\newblock {\em IEEE Trans. on Microwave Theory and Techniques},
  MTT-33(10):882--893, October 1985.

\bibitem{hrg:92}
H.~J. Hrgovci\'c.
\newblock {\em J. Phys. A}, 25:1329--1350, 1991.

\bibitem{van:92}
C.~Vanneste, P.~Sebbah, and D.~Sornette.
\newblock {\em Europhys. Lett.}, 17:715, 1992.

\bibitem{vanneste:97}
S.~de~Toro~Arias and C.~Vanneste.
\newblock {\em J. Phys. I France}, 7:1071--1096, 1997.

\bibitem{mora:92}
P.~Mora.
\newblock {\em J. Stat. Phys.}, 68:591--609, 1992.

\bibitem{BC-string}
B.~Chopard.
\newblock {\em J. Phys. A}, 23:1671--1687, 1990.

\bibitem{marder:96}
M.~Marder and J.~Fineberg.
\newblock {\em Physics Today}, pages 24--29, September 1996.

\bibitem{herrmann-roux:90}
H.J. Herrmann and S.~Roux, editors.
\newblock {\em Statistical Models for the Fracture of Disordered Media}.
  North-Holland, 1990.

\bibitem{ciliberto:97}
A.~Garcimartin, A.~Guarino, and L.~Bellon anad S.~Ciliberto.
\newblock Statistical properties of fracture precursors.
\newblock {\em preprint}, 1997.

\end{thebibliography}


\end{document}